\documentclass[letterpaper,twocolumn,10pt]{z2-article}
\usepackage{usenix2019_v2,epsfig,endnotes}
\usepackage{titlesec}
\usepackage{tikz}
\usepackage{silence}
\usepackage{times}
\usepackage{datetime}

\usepackage{hyperref}
\hypersetup{
  colorlinks,
  linkcolor={red!50!black},
  citecolor={blue!50!black},
  urlcolor={blue!50!black}
}
\hypersetup{citecolor=blue,linkcolor=blue}
\usepackage{amsmath,amsopn,amssymb}
\usepackage{subcaption}
\usepackage{endnotes,microtype,xspace,graphicx,fancyvrb,multirow}
\usepackage{booktabs}
\usepackage{array,underscore,relsize}
\usepackage[T1]{fontenc}
\usepackage{times}
\usepackage{fancyhdr}

\usepackage{listings}
\usepackage{lipsum}
\usepackage{comment}

\usepackage[normalem]{ulem}

\usepackage[labelfont=bf,font=small,belowskip=-5pt,aboveskip=3pt]{caption}
\usepackage{graphicx}
\usepackage{tabularx}
\usepackage{comment}
\usepackage{color}
\usepackage{cite}
\usepackage{fancyhdr}
\usepackage{soul}

\usepackage{amsmath}
\usepackage{cleveref}
\usepackage[ruled,linesnumbered]{algorithm2e}

\titlespacing*{\section}{0pt}{1\baselineskip}{\baselineskip}
\usepackage[export]{adjustbox}
\titlespacing*{\section}{0pt}{1\baselineskip}{\baselineskip}
\titlespacing{\section}{1pt}{*1}{*1}
\titlespacing{\subsection}{0pt}{*0}{*0}
\titlespacing{\subsubsection}{0pt}{*0}{*0}


\definecolor{red3}{rgb}{0.80,0.00,0.00}

\newcommand{\SKQ}[1]{}
\newcommand{\SK}[1]{}
\newcommand{\SKREV}[1]{}

\renewcommand\hl[1]{#1}

\newcommand{\systemname}{\textsc{kloc}\xspace}

\def \rediskernobjpercent {\mbox{36\%}\xspace}
\def \rocksdbosmemaccessdist {\mbox{54\%}\xspace}
\def \redisosmemaccessdist {\mbox{38\%}\xspace}
\def \rocksdbprefetch {\mbox{1.27$\times$}\xspace}
\def \redisprefetch {\mbox{1.32$\times$}\xspace}
\def \sparkcontextgains {\mbox{1.6$\times$}\xspace}

\def \mrocksdbslowmem{\mbox{1.58$\times$}\xspace}

\def \mfilebenchslowmem{\mbox{2.8$\times$}\xspace}
\def \mredisslowmem{\mbox{1.8$\times$}\xspace}
\def \mparkslowmem{\mbox{1.3$\times$}\xspace}

\def \rocksnvmobjperf{\mbox{1.42$\times$}\xspace}
\def \rocksdbseqprefetch{\mbox{24\%}\xspace}

\def \RocksDBNaiveVsOptimal{\mbox{2.6$\times$}\xspace}

\def \RocksDBGainOverMigration{\mbox{1.4$\times$}\xspace}
\def \RedisDBGainOverNaive{\mbox{1.6$\times$}\xspace}

\def \eredisnetworkperf{\mbox{3.3$\times$}\xspace}

\begin{document}

\date{}

\title{Efficient Kernel Object Management for Tiered Memory Systems with KLOC\vspace{-0.2in}}
\author{
{\rm Sudarsun Kannan (Rutgers University)}, 
{\rm Yujie Ren (Rutgers University)}, 
{Abhishek Bhatacharjee (Yale University)}
}


\maketitle


\begin{abstract}
  Software-controlled heterogeneous memory systems have the potential to improve
  performance, efficiency, and cost tradeoffs in emerging systems. Delivering on
  this promise requires efficient operating system (OS) mechanisms and policies
  for data management. Unfortunately, modern OSes do not support efficient
  tiering of data between heterogeneous memories. While this problem is known
  (and is being studied) for application-level data pages, the question of how
  best to tier OS kernel objects has largely been ignored.

We show that careful kernel object management is vital to the performance of
  software-controlled tiered memory systems. We find that the state-of-the art
  OS page management research leaves considerable performance on the table by
  overlooking how best to tier, migrate, and manage kernel objects like inodes,
  dentry caches, journal blocks, network socket buffers, etc., associated with
  the filesystem and networking stack. In response, we characterize hotness,
  reuse, and liveness properties of kernel objects to develop appropriate
  tiering/migration mechanisms and policies.  We evaluate our proposal using a
  real-system emulation framework on large-scale workloads like RocksDB, Redis,
  Cassandra, and Spark, and achieve 1.4$\times$ to 4$\times$ higher throughput
  compared to prior art.
\end{abstract}

\section{Introduction}
Hardware heterogeneity is here. Vendors are coupling general-purpose CPUs with
accelerators ranging from GPUs and FGPAs to domain-specific hardware for deep
learning, signal processing, finite automata, and much
more~\cite{Becchi:FinitecFinite, Jouppi:2017:IPA:3079856.3080246,
Ankit:2019:PPU:3297858.3304049}. Memory systems are combining the best
properties of emerging technologies optimized for latency, bandwidth, capacity,
volatility, or cost. Researchers are already studying the benefits of
die-stacked DRAM \cite{Jevdjic:2013:StackedCache, Akin:DataReorgStacked3D,
Radulovic:MemwallStacked3D}, while Intel’s Knight’s Landing uses high bandwidth
multi-channel DRAM (MCDRAM) alongside DDR4 memory to achieve both high bandwidth
and high capacity \cite{knightlanding}. Non-volatile 3D XPoint memories are now
commercially available for database systems, and disaggregated memory is being
touted as a promising solution to scale capacity for blade servers
\cite{Lim:Blade}. Next-generation systems will consist of heterogeneous compute
nodes (CPU, GPU, or both) connected to multiple types of memory with different
bandwidth, latency, and capacity properties.

To harness the promise of heterogeneity, software-controlled data management is
necessary to ensure that as programs navigate different phases of execution,
each with potentially distinct working sets, data is tiered appropriately. To
optimize for performance, ideally the hottest data will be placed in the fastest
memory node (in terms of latency or bandwidth) until that node is full, the
next-hottest data will be filled into the second-fastest node up to its
capacity, and so on. As a program executes, its data must be periodically
assessed for hotness and re-organized to maximize performance. Doing this
successfully requires effective OS policies and mechanisms to determine data
reuse and control data migration.

Unfortunately, data hotness tracking and page migration in modern OSes have high
overheads and are surprisingly inefficient. Recent studies address this in many
ways, including support for transparent huge page migration~\cite{Kwon:Ingens},
concurrent migration of multiple pages and symmetric exchange of pages
\cite{Yan:Nimble}, compiler- and MemIf-based frameworks \cite{Lin:memif,
Wang:2019:Panthera} that determine data hotness via heap profiling, and machine
learning to determine page hotness \cite{Lagar-Cavilla:SDFarMem}. Other OS-based
approaches \cite{Agarwal:Thermostat, Kannan:HeteroOS} and hardware accelerated
approaches \cite{Chou:Batman, Meswani:HeteroSW, Zhao:DRAM, Jiang:CHOP} have been
proposed. Although effective, these studies focus on application-level data and
\emph{largely ignore the question of how best to tier kernel objects associated
with I/O.}

We show that ignoring kernel objects leaves considerable performance on the
table and that carefully tiering, migrating, and managing kernel objects like
inodes, dentry caches, journal blocks, network socket buffers, etc., is vital to
overall system performance. We also show -- by characterizing hotness, reuse,
and liveness properties -- that techniques previously proposed for
application-level data tiering are a poor fit for kernel objects. This is
because of three key differences between application-level pages and kernel
objects. First, kernel objects are not mapped or owned by a specific
application. Second, kernel objects can and are shared and reused across
multiple tenants. Third, kernel objects are much shorter
lived than application-level data. For these reasons, traditional hotness
scanning and page migration techniques cannot simply be extended to kernel
objects, which can neither be associated with a specific application nor are
sufficiently long-lived to tolerate the latencies of state-of-the-art
application-level page migration techniques \cite{Yan:Nimble}. 

In response, we introduce the concept of {\bf k}ernel-{\bf l}evel {\bf o}bject
{\bf c}ontexts (or {\bf{\systemname}s}). Each {\systemname} encapsulates
\SK{(groups)} a set of kernel-level objects associated with entities like files,
sockets, virtual device files etc. We build allocation, deallocation, page
placement, and page migration mechanisms and policies for {\systemname}s,
striking a compromise between tying kernel-level objects to application-level
characteristics while also adapting to the unique reuse/lifetime characteristics
of kernel-level objects. {\systemname}-based tiering allows the OS to associate
applications and kernel-level objects (thereby tracking when, how, and why
applications access kernel-level objects) while also understanding that
kernel-objects can be shared across different applications (with different I/O
behavior) and have shorter lifetimes than application-level data.
{\systemname}-based tiering also permits kernel objects to leverage aspects of
the file system and networking stacks that have already been optimized for
performance (e.g., data prefetching for I/O, etc.).  

To build {\systemname}s, we address several research challenges. Unlike
coarser-grained application-level approaches based on NUMA-affinity, which
generally tie data to particular memory sockets through large chunks of
application lifetime, files and sockets can be rapidly spawned/killed and
accessed/reused in a myriad ways (for example, RocksDB~\cite{RocksDB}, 
\SK{a widely used persistent key-value store} creates and operates on
hundreds of files through its lifetime). Some kernel objects (like file system
journals) can be shared across multiple files (and applications). Finally,
currently OSes lack support for kernel object migration entirely (kernel objects
are managed by the OS's slab allocator and remains in the location it was
allocated through time). Even worse, building kernel object migration is a
non-trivial effort -- many kernel objects (like kernel buffer pages) have
lifetimes that are so short (in tens of milliseconds), that the latency to
migrate them between memory tiers and perform associated operations like TLB
shootdowns \cite{Oskin:TLB, Meswani:HeteroSW} can be prohibitively high.

To accurately gauge the benefits of {\systemname}s, we prototype our approach in
the mainline Linux kernel v4.17. To determine the set of kernel objects per
{\systemname}, we rely on application-level system calls to identify the files
and sockets, virtual devices, etc., being accessed and the newly-allocated
kernel objects (e.g., a dentry, cache page, packet/socket buffer) that they are
associated with. Each file, socket, and a virtual device has its own
{\systemname}. We also associate kernel objects with CPUs in order to maintain
information about CPU- and application-wide {\systemname} association. To track
chains of kernel objects associated with {{\systemname}s} in an efficient
manner, we implement a lightweight object map table within the kernel. This map
is accessed by the OS to determine {\systemname}s associated with cold kernel
objects so that they can be migrated to slow memory, and is designed to obviate
the need for high-overhead page table scans to determine page hotness. We
overcome this challenge by grouping kernel objects used by files and sockets to
a large region of virtually contiguous pages. By doing this, and also then going
beyond modern OSes and implementing support for kernel object migration, we
enable good performance. To boost efficiency, we also exploit I/O stack
optimizations like prefetching of file data and insertion of prefetched cache
pages to faster memory (which is particularly beneficial for workloads with
sequential I/O access patterns). Overall, we add 6K lines of code in the Linux
kernel to implement these techniques, and make no changes to the hardware or
applications.

We quantify the benefits of {\systemname}s by using a two-socket system to
emulate a two-tier memory system with a fast and slow memory, similar to prior
work \cite{Yan:Nimble, Linux:Slab, Hanson:Throttling, Kannan:pVM}. We perform
end-to-end evaluations with RocksDB (a persistent key-value store), Redis (a
network-intensive memory store), Cassandra (a distributed wide column store),
and Spark (a distributed general-purpose cluster-computing framework), in
addition to microbenchmarks like Filebench. Our \hl{performance gains of} 
up to 1.4$\times$ and 4$\times$ respectively show the
performance potential of {\systemname}s, and lay the groundwork for further
exploration of kernel object tiering mechanisms and policies in the systems
research community.

\section{Background and Related Work}
\label{sec:background}
Advances in heterogeneous memory hardware have motivated the need for
efficient management of memory resources that vary in capacity, speed, and cost.
We first discuss hardware and software trends, followed by related work on techniques for memory heterogeneity management and their
limitations.

\subsection{Heterogeneous Memory Trends}
Several heterogeneous memory technologies, such as non-volatile memory (NVM),
Hybrid Memory Cube (HMC), and High Bandwidth Memory (HBM) will coexist with
traditional DRAMs.  On-chip memory such as stacked 3D-DRAM,
Hybrid Memory Cube (HMC) and High Bandwidth Memory
(HBM)~\cite{Chou:Batman,Meswani:HeteroSW} are expected to provide 10$\times$ higher
bandwidth and 1.5$\times$ lower latency, but provide
8-16$\times$~\cite{Akin:DataReorgStacked3D, Black:Intel3D, AMDHBM, Oskin:TLB} lower capacity
compared to DRAM. Other technologies, like NVMs, offer 4-8$\times$ higher capacity compared to DRAMs but
suffer from 2-3$\times$ higher read latency, 5$\times$ higher write latency, and 3-5$\times$ lower random
access bandwidth compared to DRAM. Several file systems
~\cite{Dulloor:PMFS, NOVA, Condit:BPFS, Wu:SCMFS} and user-level
libraries~\cite{Intel:NVML, Volos:Mneymosyne} have been proposed to exploit
persistence, as have approaches to \hl{integrate them} as virtual
memory~\cite{Dulloor:xmem, Kannan:pVM}. Given the differences in bandwidth, latency, and
capacity, heterogeneous memory systems will increase the complexity of the memory management
software stack.

\subsection{Managing Heterogeneous Tiered Memories}

\par \noindent{\bf Hardware-level management.}
There have been several prior proposals to manage memory heterogeneity in the
hardware. Batman modifies the memory controller to randomize
data placement for increasing the cumulative DRAM and stacked 3D-DRAM bandwidth~\cite{Chou:Batman}.
Meswani et al.~\cite{Meswani:HeteroSW} discuss extending the TLB and the memory
controller with additional logic for identifying page hotness. To reduce page
migration cost, Dong et al.~\cite{XDong:HeteroController} propose SSD FTL-like
mapping of physical
addresses dynamically~\cite{Chung:FLTsurvey}. Oskin et al. propose an architectural
mechanism to selectively invalidate entries in the TLB for reducing the TLB
shoot-downs during migrations~\cite{Oskin:TLB}. Ramos et al. propose a
hybrid design with hardware-driven page placement policy and the OS periodically
updating its page tables using the information from the memory controller~\cite{Ramos:NVMPagePlacement}.

\vspace{2mm}
\par \noindent{\bf Software-level management.}
Several recent studies augment traditional OS approaches to track page hotness
by scanning page tables to migrate application pages of different sizes
\cite{Oskin:TLB, Gupta:HetroVisor, Yan:Nimble, Lagar-Cavilla:SDFarMem,
Lin:memif, Kannan:HeteroOS, Agarwal:Thermostat}. These approaches extend work
originally proposed by Denning~\cite{Denning:WorkingSet} for disk swapping.
Gupta et al.  propose HeteroVisor~\cite{Gupta:HetroVisor}, which uses page
hotness tracking and migration techniques for virtualized datacenters, whereas
Kannan et al.~\cite{Kannan:HeteroOS} propose on-demand data placement for
virtualized datacenters. Yan et al.~\cite{Yan:Nimble} proposes techniques to
accelerate page migration in heterogeneous memory systems by increasing
parallelism. Lagar-Cavilla et al~\cite{Lagar-Cavilla:SDFarMem} propose a
combination of OS-level hotness scanning combined with machine learning for data
placement across fast and slow memories. Many of these techniques extend the
concept of NUMA-affinity to data pages. That is, they associate applications to
particular memory sockets in order to accelerate memory access from CPUs that
are physically closer.

\vspace{2mm}
While these approaches are beneficial for application-level data, kernel object
management for heterogeneous memory remains unexplored and is in its infancy
(for example, there is no support for kernel object migration in modern OSes).

\begin{table}[ht!]
\centering
\footnotesize
\begin{tabular}{|p{1.5cm}|p{4.5cm}|p{1.5cm}|}
\hline
  {\bf Application} & {\bf Description} & {\bf \hl{Resident Mem. Size}} \\ \hline
RocksDB~\cite{RocksDB} & Facebook's persistent key-value store based on
  log-structured merge tree;  Workload: Widely-used
  DBbench~\cite{LevelDB:Googgle} with 1M keys. & 8.4GB \\ \hline
Redis~\cite{Redis} & Network-intensive key-value store with support for
  persistence; uses \emph{Redis Bench}, 4 millions ops., 75\%/25\% Set/Get distribution. & 14GB \\ \hline
  Filebench~\cite{Filebench} & File system benchmark; uses eight threads, 8GB
  per-thread, performing sequential and random reads. & 16.3GB \\ \hline
  Cassandra~\cite{Cassandra} & Java-based NoSQL DB; run with
  YCSB~\cite{Cooper:YCSB} workload using eight threads, 50\% read-write ratio. & 11GB \\ \hline
  Spark~\cite{Zaharia:Spark} & Apache Spark; performs Terrasort on 20GB data
  using sixteen threads and uses Hadoop file system. & 32.1GB \\ \hline
\end{tabular}
  \caption{\textbf{\hl{Applications and workloads.}}\hl{Table also shows application resident set size (in GB).}}
\label{tab:applications}
\end{table}

\begin{table}
\centering
\footnotesize
\begin{tabular}{|p{1.7cm} p{6cm}|}
\hline
\hline
 \multicolumn{2}{|c|}{\bf Experimental Environment} \\
 \hline
Processors & 2.4 GHz Intel E5–2650v4 (Broadwell),
20 cores/socket, 2 threads/core\\ \hline
Cache  &  512~KB L2, 25~MB LLC \\ \hline
Memory Sockets & Two 80~GB sockets configured as NUMA nodes, max bandwidth of \SKREV{30~GB/sec}\\ \hline
Storage & 512~GB NVMe with 1.2~GB and 412~MB sequential and random access bandwidth\\ \hline
OS & Debian Trusty — Linux v4.17.0 \\ \hline
\end{tabular}
  \caption{\textbf{System configurations.}}
\label{tab:sysconfigurations}
\end{table}

\section{Characterizing Reuse, Liveness, and Access Properties of Kernel-Level Objects}
\label{sec:motivate}
\hl{For optimal application performance in heterogeneous memory systems, placing not only 
application-level pages but also kernel pages to faster memory is critical. However page placement of kernel pages (and objects) is not well studied.}
To understand the need for kernel object placement in heterogeneous memory
systems, we next analyze real-world I/O-intensive applications.

\subsection{Experimental Methodology}
Quantifying the performance of kernel object tiering requires a
platform that permits end-to-end execution of large scale workloads (we use
those summarized in Table \ref{tab:applications}) with full-system effects.
Cycle-accurate simulators are too slow and lack the detail necessary to (easily
and accurately) study kernel-level structures in the virtual memory, storage,
and network subsystems \cite{Chou:Batman, Meswani:HeteroSW,
XDong:HeteroController}. Ideally, we would use a commercially-available
heterogeneous memory platform with support for flexible tiering of kernel
objects. Regrettably, there are no commercial platforms that can be configured
to do this yet. For example, we considered Intel's DC memory
\cite{intel:3dxpoint}, which attaches a persistent Optane memory side-by-side
with DRAM. Unfortunately, this system can currently only be configured such that
the persistent memory is a direct-access file system accessible via custom
user-level runtimes, or the DRAM is a direct-mapped L4 cache of the persistent
memory. There is \hl{no way to configure} the DC memory platform to make it
entirely visible (or kernel objects visible) to the virtual memory sub-system,
which is what we need for our studies.

Therefore, while we will revisit Intel's DC platform in Section~\ref{sec:eval}, 
and for now we use a two-socket memory system to emulate a two-level tiered memory in a manner that is similar to recent work \cite{Yan:Nimble, Oskin:TLB,
Gupta:HetroVisor, Hanson:Throttling, Kannan:pVM}. These sockets have the
architectural configuration described in Table \ref{tab:sysconfigurations}. Like
previous work \cite{Hanson:Throttling, Gupta:HetroVisor, Kannan:pVM}, we emulate
a fast memory on one of the sockets, and a slow memory on the other by applying
thermal throttling to slow down the latter. The slow memory's bandwidth and
latency are configured by modifying the PCI-based thermal registers. The
flexibility of this platform enables exploration of a generic
software-controlled heterogeneous framework as we can vary capacities of the
memory nodes, as well as their latency/bandwidth characteristics. For our
studies, we vary capacity/bandwidth differences between the fast and slow 
memories from 2$\times$ to 16$\times$. Furthermore, for some of our experiments,
we controlled application/kernel page placement in the fast/slow memories by
adding hooks in Linux's memory management stack to redirect page allocations.
All workloads are executed on the 20 CPUs of the node associated with fast
memory. We will publicly release our kernel and tools so that they can be used
by the community for follow-up studies.

\subsection{Experimental Results}
Table \ref{tab:applications} shows that we focus on large-scale workloads that
are compute-, file-, and network-intensive in order to stress-test our approach.
We also use Filebench, which is a mixed random write and read access workload.
We structure our studies around the following questions:

\begin{figure*}[ht!]
\vspace{-0.2in}
\begin{subfigure}{.30\textwidth}
  \centering
  \includegraphics[width=.90\linewidth]{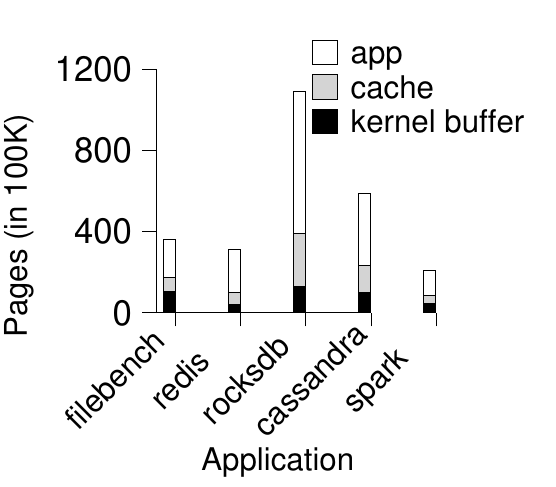}
    \caption{\textbf{Page allocation distribution.} 
    \footnotesize{The bars show distribution across heap, page cache
    (OS), and kernel buffers (slab pages). \hl{The y-axis show overall pages allocations during an application's lifetime.}}}
  \label{fig:m-pagecache}
\end{subfigure}\hfill
\begin{subfigure}{.30\textwidth}
  \centering
\includegraphics[width=.90\linewidth]{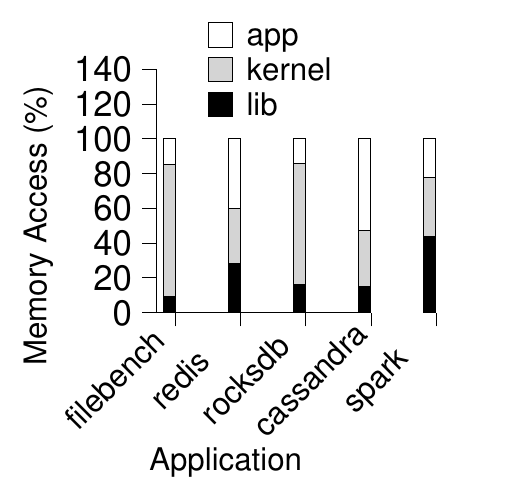}
 \vspace{-0.1in}
  \caption{\textbf{Memory access distribution.} 
  \footnotesize{The bars show  memory access distribution (in \%) across App, Kernel (OS), and other user-space libraries.}}
\label{fig:m-accessdistribution}
\end{subfigure}\hfill
\begin{subfigure}{.30\textwidth}
  \centering
  \vspace{-0.1in}
  \includegraphics[width=.90\linewidth]{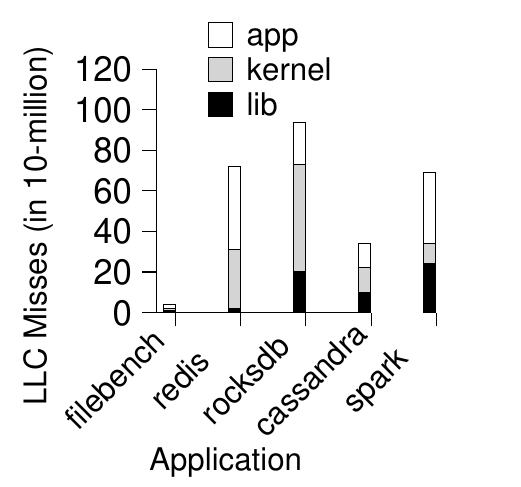}
  \caption{\textbf{Last-level cache miss distribution.}} 
 \label{fig:m-llcmissdistribution}
\end{subfigure}
\vspace{-0.1in}
\caption{\bf Memory allocation, access, and last-level cache miss distribution.}
\vspace{-0.2in}
\end{figure*}

\vspace{1mm}
\noindent \textbf{How are kernel memory objects allocated and accessed?} Modern
data center applications are known to be I/O intensive and expend considerable
portions of their runtime within the kernel-level file system and networking
stacks. For example, 40\% of the runtime of RocksDB is spent within the file
system code path. What is less well-known, however, is that these applications
also allocate millions of memory pages for kernel objects that perform I/O
caching, in-memory metadata and book-keeping structures (like radix trees for
caches), journals and logs, as well as ingress and egress network socket
buffers.

Figure~\ref{fig:m-pagecache} quantifies the number of pages allocated for
different kernel objects. The data shows that all the
workloads allocate many page cache pages and kernel buffers (using slab
allocations via {\it kmalloc}). Filebench uses eight I/O threads to
simultaneously write 4KB blocks to separate files. 
The write and read operation entails allocation of page cache (for writing data
or bringing data from disk) as well as updates to system metadata structures,
which involves allocating journals, radix tree, block driver buffers, etc. 
Consequently, both page cache and kernel metadata
allocations increase significantly compared to user-level pages. In contrast,
RocksDB updates hundreds of 4MB files with key-value data. Therefore, slab
allocations for inodes, dentry caches, radix tree nodes (for the indexing
cache), driver block I/O and journals are all frequent and contribute to
{\rediskernobjpercent} of the pages allocated to kernel objects. Redis is
network-intensive and allocates many pages for ingress and egress socket
buffers, and also page cache pages to periodically checkpoint key-value store
state to a large file on disk \cite{Redis}. Spark~\cite{Zaharia:Spark} uses the
Hadoop file system (HDFS) to store and checkpoint data
(RDDs~\cite{Zaharia:Spark}). Note that HDFS is run as a separate process. HDFS
maintains user-level cache and periodically updates page cache (so less kernel
buffer pages)

We also profile the frequency with which these different kernel objects are accessed in Figure~\ref{fig:m-accessdistribution} and the distribution of last-level cache misses in Figure~\ref{fig:m-llcmissdistribution}. Even though fewer kernel buffers are
allocated (see Figure \ref{fig:m-pagecache}), they are accessed more often
than other kernel objects. To understand why, consider, for example, a file
write in Filebench. The virtual file system (VFS) looks up the page cache radix
tree, allocates a new page if the necessary, inserts the page into the radix
tree, performs metadata/data journalling with logging, and finally, commits to
storage. These steps are more memory-intensive than writing the data to the page
cache. In fact, scaling the workload inputs leads to a sharp
increase in LLC misses due to higher traffic to kernel buffers. 
 \hl{Filebench's spends 86\% of execution time inside the OS, and hence, the memory
 accesses increase proportionately}, compared to RocksDB
 (\rocksdbosmemaccessdist) and Redis (\redisosmemaccessdist).

\begin{figure}[t]
\centering
\vspace{-0.3in}
\includegraphics[width=.90\linewidth]{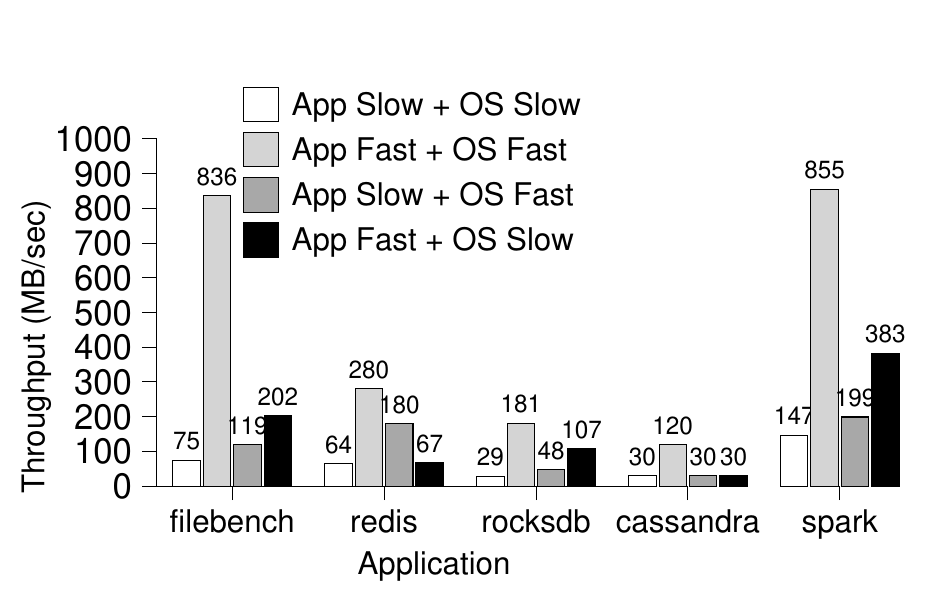}
  \vspace{-0.1in}
  \caption{{\bf Impact of kernel object (page) placement.} 
  \footnotesize{Fast memory capacity is set to 1/8th (5GB) capacity of 
  Slow memory (40GB), and Slow Memory bandwidth is set to 1/5th (4GB/sec) of Fast
  Memory.}}
 \label{fig:m-osimpact}
\end{figure}

\begin{figure}[t!]
  \centering
 \includegraphics[width=.90\linewidth]{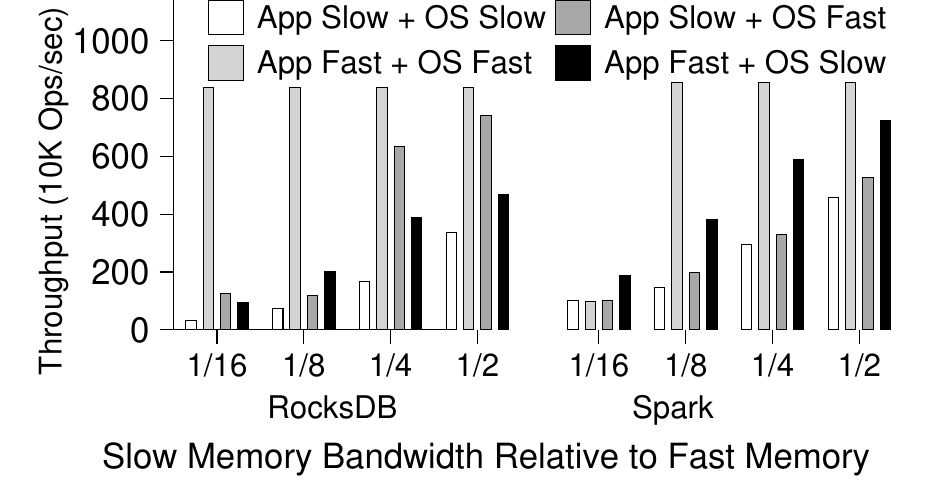}
 \caption{{\bf Memory bandwidth sensitivity.} \footnotesize{The x-axis varies slow memory bandwidth relative
 to fast Memory, and the y-axis shows throughput.}}
  \label{fig:m-osimpact-sensitive-BW}%
\end{figure}

\begin{figure}[t!]
  \centering
 \includegraphics[width=.90\linewidth]{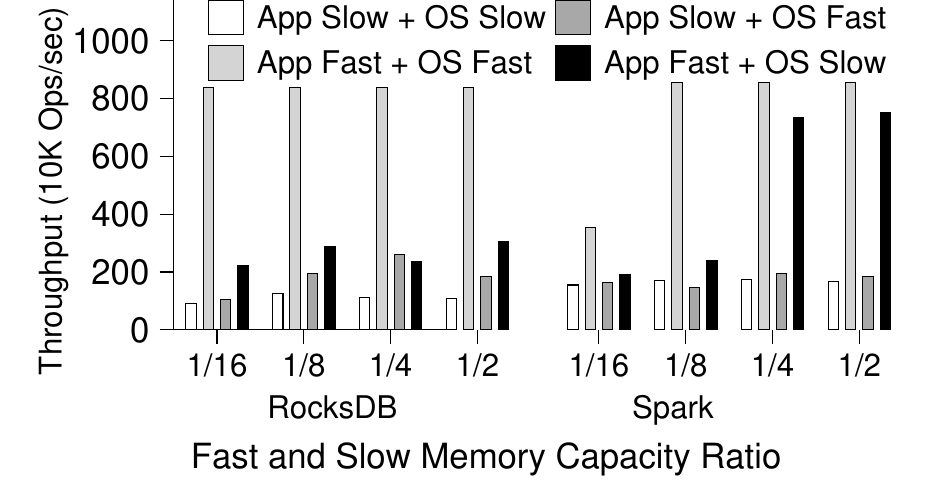}
 \caption{{\bf Sensitivity to fast memory capacity.} 
 \footnotesize{The x-axis show fast memory capacity ratio
 relative to an all-fast Memory system, and the y-axis shows throughput
 impact. App Fast-OS Fast shows optimal case performance in an all-fast
 memory system. The slow Memory bandwidth is set of 1/16th of Fast Memory}}
  \label{fig:m-osimpact-sensitive-cap}%
\end{figure}

\vspace{2mm}
\noindent \textbf{How does tiering of kernel objects impact performance?} To
study the impact of kernel object placement in fast/slow memory, we configure
the capacity of the fast memory so that it cannot fit all the application's
user-level and kernel-level pages. Our results, illustrated in
Figure~\ref{fig:m-osimpact}, assume that fast and slow memory are 5GB and 40GB
respectively, and that slow memory has a bandwidth of 5GB/second, thereby
emulating a 5$\times$ bandwidth difference between fast and slow memory. This is
similar to recent work~\cite{Meswani:HeteroSW, Gupta:HetroVisor,
Kannan:HeteroOS} and is representative of the bandwidth differences 
between HBM and NVM technologies relative to DRAM. In tandem,
Figure~\ref{fig:m-osimpact-sensitive-BW} and
Figure~\ref{fig:m-osimpact-sensitive-cap} show the impact of varying slow
memory bandwidth and fast memory capacity. The \emph{App Slow + OS Slow} bars show
the worst-case scenario where all pages are placed in slower memory, \emph{App
Slow + OS Fast} shows the case where only the kernel pages are placed in fast
memory,~\emph{App Fast + OS Slow} shows the case where only application-level
pages are placed in fast memory, and \emph{App Fast + OS Fast} shows an ideal
case all pages fit in fast memory. The y-axis shows the normalized throughput.

Placing both application and kernel pages (\textit{App Slow + OS Slow}) in slow
memory degrades performance across all workloads. As shown in
Figure~\ref{fig:m-osimpact}, placing only kernel pages
(\textit{App Slow + OS Fast}) to limited-capacity fast memory does improve
performance; for example, RocksDB and filebench improve by~\mrocksdbslowmem
and ~\mfilebenchslowmem respectively compared to~\textit{App Slow + OS Slow}. In
real-world settings, one would not just place kernel pages in fast memory (but
would also do so for application-pages as much as possible) but this experiment
shows that even just tiering kernel objects appropriately impacts performance.
For network (and storage) intensive Redis, placing kernel pages in fast memory boosts
performance by \mredisslowmem over~\textit{App Slow + OS Slow}. 
In Spark, we note a high contention between Spark compute pages (heap) 
and HDFS storage (kernel pages) for a limited-capacity fast memory. 
Only placing kernel pages in fast memory improves performance by 
\mparkslowmem, mostly due to page cache placement in fast memory.

Finally, suppose we place application pages in fast
memory but prevent kernel objects from being in fast memory. While Redis
improves marginally, this is not the case for Filebench, which spends 86\% of
execution inside the OS.

Figure~\ref{fig:m-osimpact-sensitive-BW} and
Figure~\ref{fig:m-osimpact-sensitive-cap}  show the sensitivity of RocksDB
(highly I/O and OS-intensive) and Spark (mostly compute-intensive with
intermittent I/O) towards lowering memory bandwidth or reducing fast memory
capacity.  The x-axis in Figure~\ref{fig:m-osimpact-sensitive-BW} shows the
increasing ratio of slow memory bandwidth to fast memory bandwidth,  whereas
Figure~\ref{fig:m-osimpact-sensitive-cap} shows the increasing fast memory
capacity ratio. The results show that reducing fast memory capacity or lowering
slow memory impacts both application and the impact of placing kernel objects to
slower memory affecting RocksDB significantly.  Overall, these results show that
the choice of where to place kernel objects impacts performance substantially
and that there is consequently a need to go beyond prior work and devise
efficient tiering of kernel objects.

\begin{figure}[t]
\centering
  \includegraphics[width=6cm,height=3.5cm]{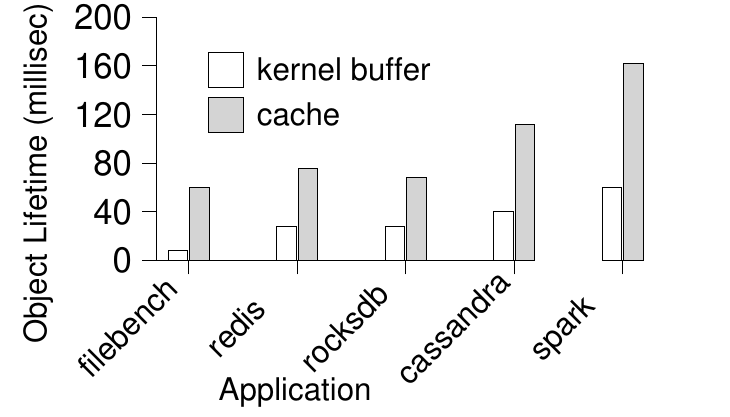}
   \vspace{-0.05in}
  \caption{{\bf Active lifetime and access interval. } {\footnotesize Bars show average active life time of cache and kernel-buffer pages before recycled; y-axis in milliseconds.}} 
\vspace{-0.1in}
\label{fig:m-ospagelife}
\end{figure}

\noindent \textbf{What is the lifetime of a typical kernel object?}
Figure~\ref{fig:m-ospagelife} shows the lifetime of OS cache and kernel buffer
(slab) pages. Conceptually, kernel buffer pages expire after they are freed.
Cache pages remain until they are evicted from memory pressure (we mark them
as expired when they are added to the LRU list).

Figure~\ref{fig:m-ospagelife} shows that RocksDB and Redis have cache pages with
average lifetimes of less than 160 milliseconds, and smaller kernel pages are
even shorter-lived, at 60 milliseconds. To understand why kernel objects are
short-lived, consider, for example, a file write. A page cache page is
allocated, \hl{the user data is copied to the cache page}, a radix tree node is
allocated using the slab allocator~\cite{Linux:CacheRadixTree}, and the cache
page is inserted. The page cache page remains inactive until subsequent
reads/writes or commits (i.e., \textit{fsync()}) to the disk. In contrast,
kernel buffers such as radix tree nodes are frequently queried, allocated, and
deleted due to tree rebalancing or cache page deletion. Other in-memory
structures such as dentry caches and in-memory journals are also frequently
allocated and deleted when data and metadata are updated.

These observations showcase the limitations of prior work like
Thermostat~\cite{Agarwal:Thermostat}, which uses a 30-second interval between
two hotness tracking iterations. This relatively large time period was used to
because scans of page tables to ascertain hotness and
invalidate TLB entries are long latency events. Our results show, however, that kernel objects are far
too short-lived to be amenable to such high intervals.

\section{Our Approach: KLOC-Based Tiering}
\label{sec:principles}
Having quantified the performance challenges posed and opportunities offered by
tiering of kernel objects in generic software-controlled heterogeneous memory
systems, we consider how to go beyond prior work (which neglects kernel object
tiering) and devise kernel tiering. Our goals are high performance and ready
implementation in commercial OSes.

At first blush, one might consider extending OS support for NUMA affinity to
also include kernel objects. This approach would permit placement of kernel
objects -- just like application pages -- in memory devices in a manner that
attempts to minimize distance between CPUs and the data they frequently access.
Unfortunately, such NUMA affinity approaches are not viable for kernel objects,
which can be shared/reused across applications (making it challenging to assign
a single affinity to kernel objects) and have lifetimes much shorter than
application pages (making existing ways of measuring hotness and migrating pages
inapplicable). 
Current OSes, which includes Linux, FreeBSD, Solaris, lack capability to
associate kernel objects with application entities or provide fine-grained 
of placement kernel objects.
Instead, we use the concept of {\systemname}s to encapsulate
groups of kernel objects into logical entities -- associated with files and
network sockets -- that can be managed together in a lightweight manner. Using
these entities as a unit of movement enables finer-grained decision-making about
allocations, placements, and migration of kernel objects associated with an
entity (i.e., a \systemname), than using NUMA affinity. These benefits are
crucial for good performance but does require extending existing filesystems,
network stacks, the OS slab allocator, journals, and device drivers with support
for {\systemname}s. A key feature of our approach is to use tap into
application-level system calls to accurately track {\systemname}s, and kernel
level maps that lets us identify hot kernel objects much faster than traditional
approaches that scan the page table. This in turn permits us to migrate kernel
objects fast enough so that its benefits are not outweighed by relatively short
kernel object lifetimes. Finally, we leverage existing and highly-optimized I/O
stack optimizations such as adaptive I/O prefetching (also known as readahead)
and techniques to speculatively place I/O cache and filesystem objects to fast
memory for further boost performance.

\section{KLOC Design}
\label{sec:design}
We next discuss the key design and implementation details of {\systemname}.  We
base our design discussions on support for the file system and network stacks,
followed by ways to support effective kernel object placement and migration and
also exploit traditional OS optimizations such as I/O prefetching.

\begin{figure}[t]
\centering
 \vspace{-0.2in}
\includegraphics[width=8.5cm,height=4.5cm]{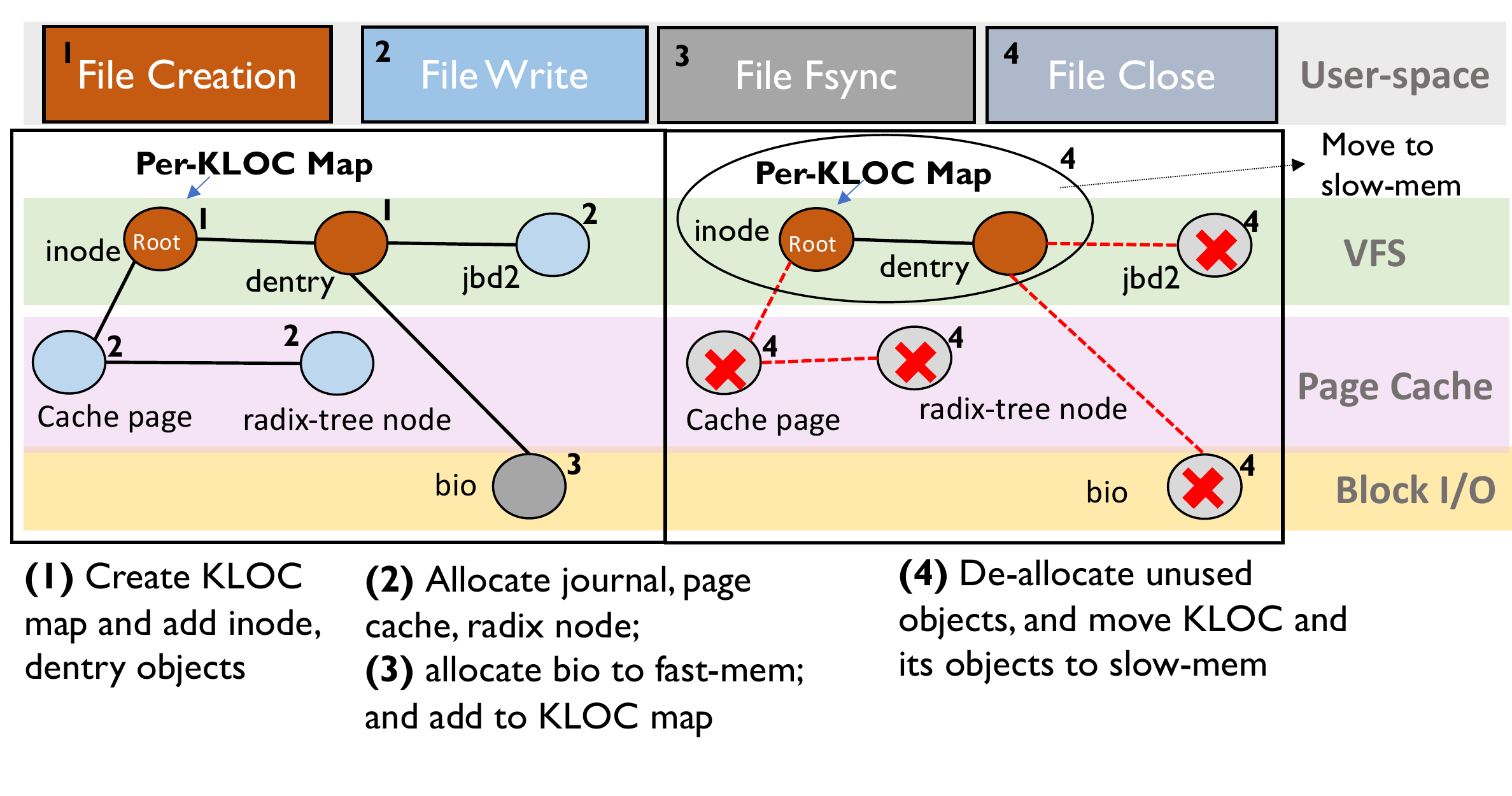}
 \vspace{-0.05in}
 \caption{{\bf {\systemname} support for storage Stack.} {\normalfont
 \footnotesize{Creation of a {\systemname} map and addition of VFS, file system,
 and device driver kernel objects.}}}
 \vspace{-0.15in}
\label{fig:design}
\end{figure}

\subsection{KLOC Placement Policy} 
We start with a set of page placement constraints. First, kernel objects are
short-lived, and immediate placement of currently active objects to faster
memory is critical. Second, because faster memory is usually lower-capacity,
migrating inactive objects to slower memory and making way for active objects is
critical. Third, reducing the frequency of long-latency hotness scans and
reducing migration overheads is critical. Finally, application pages are always
prioritized to use faster memory unless their pages become inactive.
Importantly, {\systemname} object grouping, placement, and migration are not
tightly bound to a specific placement policy.

\subsection{KLOCs for File Systems}
\label{d:fs}
The workloads that we use spend a significant fraction of execution time in the
file system dealing with page caches, in-memory structures like inodes, dentry
caches, radix trees, block-device buffers, and journals, and in the network
subsystem where they interact with ingress/egress socket buffers and network I/O
queues.

\begin{figure}[ht!]
\centering
 \vspace{-0.1in}
 \includegraphics[width=8cm,height=5.5cm]{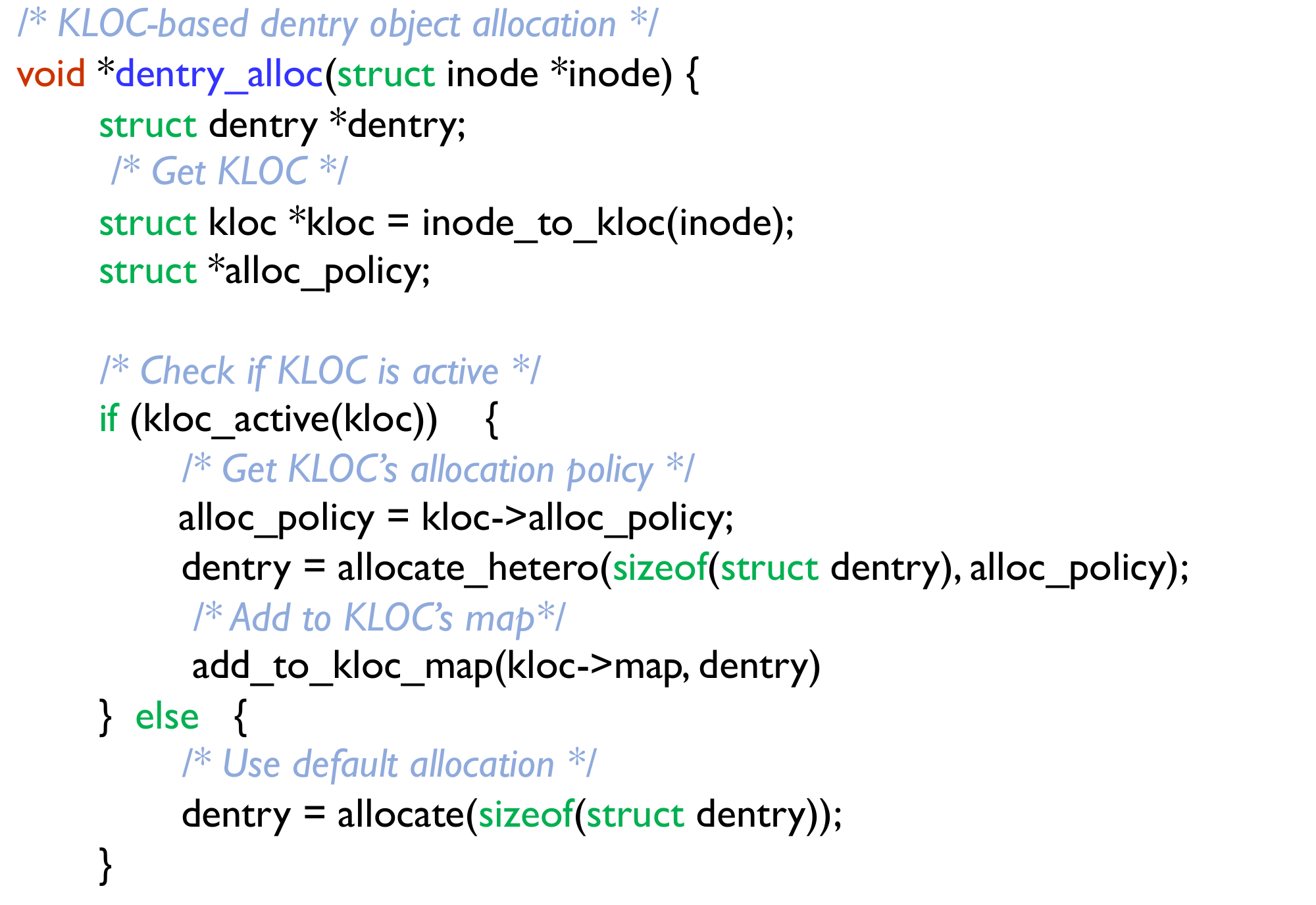}
  \caption{{\bf Pseudocode for {\systemname}-based dentry object allocation and
  mapping. }{\footnotesize{Dentry is used to keep track of the hierarchy of files in
  directories. The pseudocode code checks if a file-based {\systemname}
  (represented as inode) is active, allocates the dentry based on
  {\systemname}'s allocation policy, and finally adds the dentry object to
  {\systemname} map.}}}
 \label{c:list-add}
 \vspace{-0.2in}
\end{figure}
To the best of our knowledge, current OSes lack abstractions and the capability
to group kernel objects and efficiently \hl{place them in} heterogeneous memory
systems. 
Our focus is on CPU interactions for these workloads but as accelerators
like GPUs increasingly offer access to system services \cite{Vesely:GPYSYS}, we
expect {\systemname}s to be also useful for other processing paradigms beyond
CPUs.

\subsubsection{\textbf{Associating KLOCs with Kernel Objects}}
Ideally, {\systemname}s should be associated with entities that are accessible
(and meaningful) to both applications and the OS and allow sharing of kernel
objects across applications when required.  
Consider, for example, the notion of a {\it file}, which is visible to both the
application and the OS code paths responsible for kernel object allocation and
I/O servicing. During a file create operation, a set of in-memory kernel objects
such as inode, dentry cache, and journal blocks are allocated. During a file write
operation, the virtual file system (VFS) allocates cache pages, radix
tree nodes for managing the cache pages, journal records (for
crash-consistency), and extents~\cite{F2FS}. When the block driver commits
in-memory pages of a file to disk, the block driver allocates a file's block I/O
structures. All these allocations, whether initiated by the application or OS,
are associated with the notion of a file, making it the appropriate entity
around which to group the kernel objects allocated to it. We therefore use a
file's inode used across all layers of file system to create a {\systemname} to track all associated kernel objects, (i.e., the VFS, the file system, and the device driver).


\subsubsection{\textbf{Managing File System KLOCs}}
We create a map structure in the inode {\systemname} for tracking all 
kernel objects associated with each file. The map structure is implemented as
red-black tree and maintains kernel objects in the virtual file system layer
(VFS), the dentry cache (used for name lookup), the page cache, the actual file
system (without loss of generality, we use Ext4 file system in our implementation), the log
manager (for crash-consistency), and the block device driver. We show an
example of this map in Figure~\ref{fig:design}. 

One of the key challenges in maintaining inode {\systemname}s is to understand
when to create and update kernel objects. In order to do this, we use OS system
calls such as \textit{create(), write(), read(), close()} as semantic hints.
Figure \ref{fig:design} shows that during file creation, a new inode and its map
is created (with the inode as root), and the dentry object is added. On a file
write operation, cache pages, their radix tree nodes, and the journal records
(JBD2) are added to the map too. After a file is closed, the cache entries are
removed. Consequently, the {\systemname} is deleted only when the file and inode
are deleted \cite{Kannan:DevFS}.

\subsubsection{\textbf{Per-CPU Kernel Object Grouping}}
Modern applications run on tens of CPUs and future applications may run on hundreds of CPUs sharing and access per-thread files. Examples of such applications include those that perform graph computations and maintain key-value stores. In such cases, it is beneficial to be able to manage {\systemname}s across applications and CPUs. 

Therefore, our approach assumes that a file (i.e., inode) becomes active 
when thread(s) perform I/O on it. To integrate a notion of CPU numbers with
{\systemname}s, we exploit Linux's per-CPU data structures. Each thread (or CPU)
is represented by a task structure (\textit{task\_struct}) that represents the
active process currently scheduled on the CPU.  We extend
the~\textit{task\_struct} with an \emph{active context} state that represents
the files actively being accessed by the CPU. We leverage file system I/O system
calls, such as open, read, write, close, and others to identify the kernel
objects being accessed by each CPU. These kernel objects are added to the
per-CPU context map, as shown in the the pseudocode in  Figure~\ref{c:list-add}.
This approach enables {\systemname} to group kernel objects concurrently
accessed by multiple CPUs.

\subsection{KLOCs for the Networking Subsystem. }
\label{d:nw}
To group kernel objects allocated by network subsystem, we use socket
descriptors. Within Linux (and other OSes), these are also implemented as
inodes, making our file system {\systemname} design easily extensible to
networking. Sockets present an appropriate entity for {\systemname} creation as
they are accessible by both the application and the OS. Page placement and
migration policies are applied to pages grouped with respect to sockets.

\subsubsection{\textbf{Associating KLOCs with Kernel Objects}} The network stack
uses packet buffers (\emph{skbuff}) to send and receive application data. The
packet buffers allocated in the ingress and egress path dominate kernel object
allocations. The ingress and egress path in the network stack comprises of
multiple layers, including TCP, UDP, IP, and network device driver (e.g., NAPI).
To reduce overheads of copying network packets across these layers, user-level
buffers are copied to socket buffers and reused across other lower layers (TCP,
UDP, IP, and NAPI) before being copied over to the NIC. For the receive path,
the device driver is responsible for allocations, which are subsequently reused
by upper layers of the networking stack. In addition to the socket buffers,
kernel objects such as network queues are also allocated, but constitute a small
fraction allocated memory. Our networking {\systemname}s group these per-socket
network kernel objects, enabling better memory placement and migration.

As with inode {\systemname}s, socket {\systemname}s also use a map structure to
track kernel objects associated with a socket. Therefore, {\systemname} uses the
system calls responsible for socket creation (\textit{socket(), open()}) to
initiate per-socket map structures. When a CPU invokes an egress
(\textit{send()}), ingress (\textit{recv()}), and polls, {\systemname} marks the
socket active and adds network kernel objects to the per-socket map structure.
Whether newly-allocated kernel objects are allocated to fast or slow memory
depends on how the socket is managed. In general, all kernel objects of an
active socket {\systemname} are directed to fast memory. 

\subsubsection{\textbf{Ingress and Egress Paths}} The egress path of a network
stack is generally synchronous, and the network stack objects, including the
socket buffers, are allocated during \emph{send()} operations. As a result,
grouping socket buffers and network stack kernel objects in the egress path
involves simply adding them to the socket's context map structure. Unlike the
egress path, the ingress (network receive) path is dependent upon the
asynchronous arrival of packets. As network packets arrive, the device driver
process inside the OS allocates a generic packet buffer but does not know the
socket information to which this packet belongs. This information is extracted
in a higher layer of the TCP stack. This creates a research challenge -- \hl{how to 
group egress packet buffers to a socket} {\systemname} as early as possible,
so as to apply the appropriate set of memory tiering policies to them. 

An initial, straightforward approach is to extract the packet's entire header
and identify the associated socket number within the driver itself (before
transfer of control to the higher TCP layers), add to the per-socket context
map, and apply memory allocation policies of the socket context. Unfortunately,
this naive approach means that we inspect the packet for socket information in
both the driver and higher layers, which is both CPU-intensive and
time-consuming. In practice, we find that the latency for these steps is
sufficiently high that they outweigh the performance benefits of placing these
buffers within faster memory. The key problem is that the buffers are so
short-lived that the additional time for header extraction becomes a performance
bottleneck.

A better approach, and the one we use, is to extend the network stack device
driver. We do indeed add code to extract socket information
within the device driver, but we improve performance by avoiding redundant work
at the higher-level layers. We do this by extending the packet buffer
structure (\emph{skbuff}) with an 8-byte socket field, which contains the socket
information extracted in the device driver. This field elides the need for
further socket information extraction at the higher levels of the TCP stack. We
also extend the device driver to add the packet to the socket {\systemname}'s
map structure and allocate the packet to faster memory (provided the socket is
active).  Extending the idea of grouping kernel objects with a socket context
deep down to the device driver provides the flexibility to group kernel objects
that are allocated asynchronously and apply uniform allocation and data
placement policies.

\subsection{Enabling Support for Kernel Object Migration}
Grouping kernel objects via inode and socket {\systemname}s and supporting
per-CPU active contexts enables better memory kernel object tiering. Unlike
modern OSes, which do not support kernel object tiering, we can allocate kernel
objects to fast memory, identify cold/inactive kernel objects, and migrate the
latter to slow memory. To build efficient techniques to identify cold/inactive
kernel objects, and build support for their migration, we go beyond prior
application-level page placement systems. {\systemname} cannot rely on
long-latency hotness scanning and page migrations (that incur~TLB
invalidations~\cite{Lagar-Cavilla:SDFarMem,Yan:Nimble}) because the kernel
objects are much too short-lived to tolerate long migration latencies.

\subsubsection{\textbf{Reducing Migration-related Costs}}
Consider page hotness scanning. With {\systemname}, when a file or socket
context is actively used, all kernel objects grouped in the context map are
placed in faster memory. If the faster memory capacity is full, the objects of
the inactive file or socket {\systemname}s (currently not accessed by any CPU)
are moved from faster to slower memory. Migrating inactive {\systemname}s is
conceptually similar to the idea of garbage collecting in fast memory, without
incurring the cost of tracking each and every kernel object. We rely on quickly
identifying cold/inactive kernel objects and migrating them to slow memory. To
do this, we use and extend the Linux LRU mechanism developed initially for
swapping and adopted by prior work~\cite{Kannan:HeteroOS, Yan:Nimble}. We
introduce several extensions critical for kernel object placement in addition to
concurrent page migrations introduced by Yan et al.~\cite{Yan:Nimble}.

First, OSes such as Linux (and FreeBSD) maintain an active page list, an active
LRU list, and an inactive LRU list of pages. Due to limited faster memory
capacity, we not only migrate pages from the inactive list but also from the
active LRU list when the demand for fast memory is high. Second,  we do not wait
for Linux to identify LRU and inactive pages; instead, once a network or file
context becomes inactive, we immediately mark and migrate the pages. Third, slab
pages could be shared across one or more active and inactive file/socket
entities. To avoid undue effects, we do not migrate shared pages with active
objects. Finally, repeated migration of kernel pages between fast and slow
memory can be detrimental to performance. To avoid repeated migration, we use an
8-bit  per-page counter to track migrations and retain such pages in fast
memory. We observe a small fraction (less than 1\%) of pages that meet these
conditions due to the shorter lifetime of kernel objects.

\subsubsection{\textbf{Support for Migration of Slab Pages}}
Now consider the actual mechanism to migrate kernel objects. Traditionally,
kernel buffer pages allocated using the OS slab allocator (which excludes cache
pages or \textit{kmalloc()}) are managed independently of the application pages
and are reused across one or more applications. The slab allocator attempts to
group objects of the same type or sizes together~\cite{Linux:Slab}. Each slab
page can contain one or more kernel objects from different subsystems or shared
by different tenants. Importantly, slab kernel pages are not directly mapped to
an address space or process, and can also be accessed using a physical address. 
\emph{Current OSes do not support migration of slab pages.}

To enable migration of slab allocated kernel objects, one might consider
entirely redesigning the slab allocator. Although ideal, given the magnitude of
changes to the OS design, in {\systemname}, we propose an alternative solution
to support the migration of kernel objects. First, for kernel objects related to
files and sockets, we use \textit{vmalloc()} (virtual malloc) support inside the
OS. Using \textit{vmalloc()} provides the ability to allocate a large region of
virtually contiguous memory that also contains an anonymous address space
(anon\_vma)~\cite{Linux:vmalloc} not backable by a file. Using
\textit{vmalloc()} allows us to allocate and coalesce kernel objects of a
context to a virtually contiguous region, and use the anonymous address space to
extend Linux's page migration code to support kernel object
migration\footnote{Our current approach is limited to support the migration of
kernel objects that are not physically dereferenced.}.

\subsection{Exploiting I/O Stack Prefetching Hints}
\label{d:prefetch}
Using {\systemname}s also allows us to leverage existing OS-level optimizations such like in-memory buffering, prefetching and batching to accelerate performance. We extend Linux's \textit{adaptive readahead mechanism} (also known as I/O prefetching) to understand the notion of {\systemname}s so that workloads with temporal and spatial locality of I/O reference can be accelerated. Our approach requires no changes to I/O prefetching policies and mechanisms; we simply direct all I/O prefetches to fast memory. For example, consider that readahead speculatively reads a portion of file contents into memory with an expectation that the process working on a file will read/write that data in the future. Existing readahead mechanisms are adaptive; by tracking how often prefetched pages are actually used, the OS maintains a window of prefetch targets. When prefetches are used frequently, the window dynamically increases (to a maximum of 128MB in Linux); when the prefetches are not used often, the window is shrunk. Generally, the prefetching windows increase when the workload demonstrates sequential memory access, while it decreases when access patterns become more random~\cite{He:Contract,
LinuxReadAhead}. By placing readahead pages in faster memory, we accelerate sequential workloads. For random access workloads, the prefetch window automatically shrinks, and our approach has no punitive effective. (This also ensures that we do not over-aggressively allocate cold/inactive kernel objects to fast memory.)

\section{Evaluation}
\label{sec:eval}
Our evalulations answer the following questions:
\begin{itemize}
\item What are the benefits and implications of {\systemname}'s fine-grained placement of file system's kernel objects in heterogeneous memory systems? What is the impact on fast memory utilization?
\item How effective is {\systemname}'s support for the network subsystem?
\item Is the {\systemname}'s capability to accelerate I/O stack's prefetching optimization with fast memory pages beneficial?
\end{itemize}

\begin{table}
\centering
\footnotesize
\begin{tabular}{|p{2cm}|p{6cm}|}
\hline
Mechanisms & Description \\ \hline
 All-SlowMem & A worst-case slow memory-only system; OS and application pages
  are always placed to slow memory\\ \hline
All-FastMem & An ideal all fast memory system (best case) \\ \hline
Naive & A greedy approach that naively uses NUMA and attempts to place 
  application and OS pages to low capacity fast memory although they cannot
  fit\\ \hline 
Nimble & State-of-the-art application page placement system with concurrent migration support~\cite{Yan:Nimble}\\ \hline 
Migration-only & Migration based approach that only migrates cold pages from fast to slow memory, freeing fast memory for direct 
allocation similar to~\cite{Kannan:HeteroOS, Yan:Nimble, Agarwal:Thermostat}\\ \hline
  {\systemname}-nomigrate & {\systemname}'s kernel page placement but without migration for the
 storage stack\\ \hline
  {\systemname}-migrate-fs-noprefetch & {\systemname} with kernel page migration but
  without network stack or I/O prefetcher support (\cref{d:fs})\\ \hline
  {\systemname}-migrate-fs-nw-noprefetch & {\systemname}-migrate-fs-noprefetch with support for network stack (\cref{d:nw})\\ \hline
  {\systemname}-migrate-fs-nw-prefetch & Uses hints from the OS  I/O prefetcher
  (\cref{d:prefetch})\\ \hline
\end{tabular}
  \caption{\textbf{Evaluation Mechanisms.}}
\label{tab:sysmechanism}
\end{table}

\subsection{Methodology and Baselines}
We use a 40-core Intel Xeon 2.67 GHz dual-socket system, with 80GB memory per
socket and a 512~GB Intel Optane NVMe with a peak sequential and random
bandwidth of 1.2~GB/s and 425~MB/s respectively. We use the memory heterogeneity
emulator from ~\cref{sec:motivate} and consider a generic fast (DRAM) and slow 
(throttled) memory. As discussed in~\cref{sec:motivate} (see
Figure~\ref{fig:m-osimpact-sensitive-BW}), the fast memory capacity and slow memory
bandwidth have a direct correlation with performance. We quantify performance
sensitivity to slow memory bandwidth. We fix the fast memory capacity to 5~GB
(this is representative of recent industrial and academic
projections~\cite{Black:Intel3D, intel:3dxpoint, Gupta:HetroVisor,
Deng:Memscale, Kannan:HeteroOS}). We use the same set of applications studied
in~\ref{sec:motivate}. To understand the performance of \systemname on real NVM device (for which the bandwidth can be modified) for RocksDB, we use a 64-core, 2 TB Intel's DC platform.

\vspace{2mm}
\noindent \textbf{System Configurations.} 
We compare our approach to several other page placement and migration
approaches. These are summarized in Table~\ref{tab:sysmechanism}: (1)
\emph{All-SlowMem} represents a worst-case baseline on a system with only slow
memory.  (2) \emph{All-FastMem} represents an ideal system that can fit the
entire application workload in fast memory. (3) \emph{Naive} represents an
approach where the OS greedily places both application and kernel pages in fast
memory. When fast memory is full, subsequent allocations for application and OS
pages are served from slower memory until fast memory pages become free.  
\hl{(4) \emph{Nimble} represents the state-of-the-art system for placement of 
application-level pages. Nimble also employs concurrent page migrations}~\cite{Yan:Nimble}.
(5) \emph{Migration-only} attempts to allocate OS and application pages to fast
memory, but also identifies cold pages migrates them to slower memory similar
to~\cite{Agarwal:Thermostat}. (6) \emph{{\systemname}-nomigrate} represents a
version of our approach using {\systemname}s that group kernel objects and
allocate them to fast memory. What we omit from this is the idea of migrating
pages of inactive {\systemname}s from fast to slow memory. (7)
\emph{{\systemname}-migrate-fs-noprefetch}, which goes beyond
{\systemname}-nomigrate and also migrates cold/inactive file system kernel
objects. Finally, (8) \emph{{\systemname}-migrate-fs-nw-noprefetch} and
\emph{{\systemname}-migrate-fs-nw-prefetch} represent our final, full-blown
{\systemname} approach by also adding support for the network stack and
including I/O prefetching optimizations to fast memory.

\begin{figure}[ht!]
\begin{subfigure}{.45\textwidth}
  \centering
  \includegraphics[width=.99\linewidth]{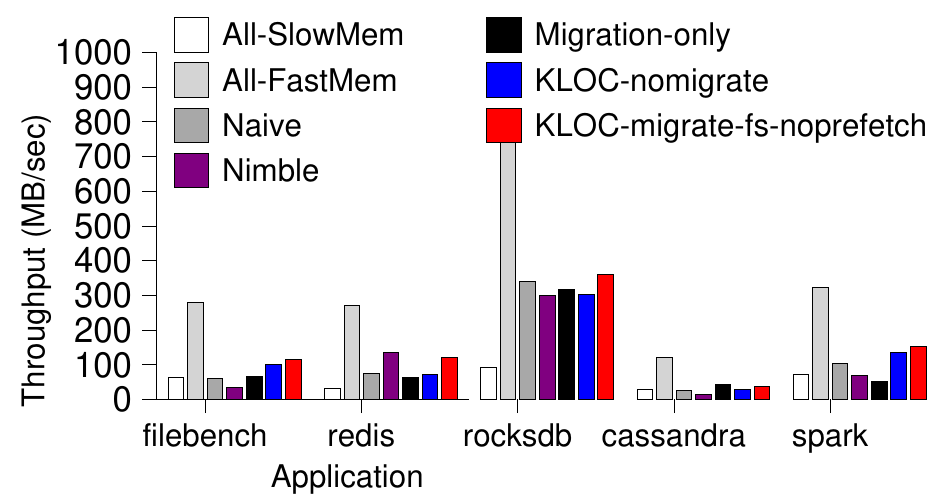}
  \captionsetup{width=0.95\linewidth}%
  \vspace{-0.2in}
  \caption{{\bf {\systemname} performance for applications and benchmarks.}
  {\footnotesize Fast memory capacity is set to 5~GB (i.e., 1/8th of
  slow memory capacity) \hl{and the slow memory bandwidth is 1/8 of the fast memory}. 
  Redis results only considers {\systemname} for file system. Y-axis shows scaled throughput (OPS/sec) for each application. \hl{Nimble shows state-of-the-art application data placement system.}}}
  \label{fig:e-allapps}
\end{subfigure}
\begin{subfigure}{.50\textwidth}
  \centering
  \includegraphics[width=.99\linewidth]{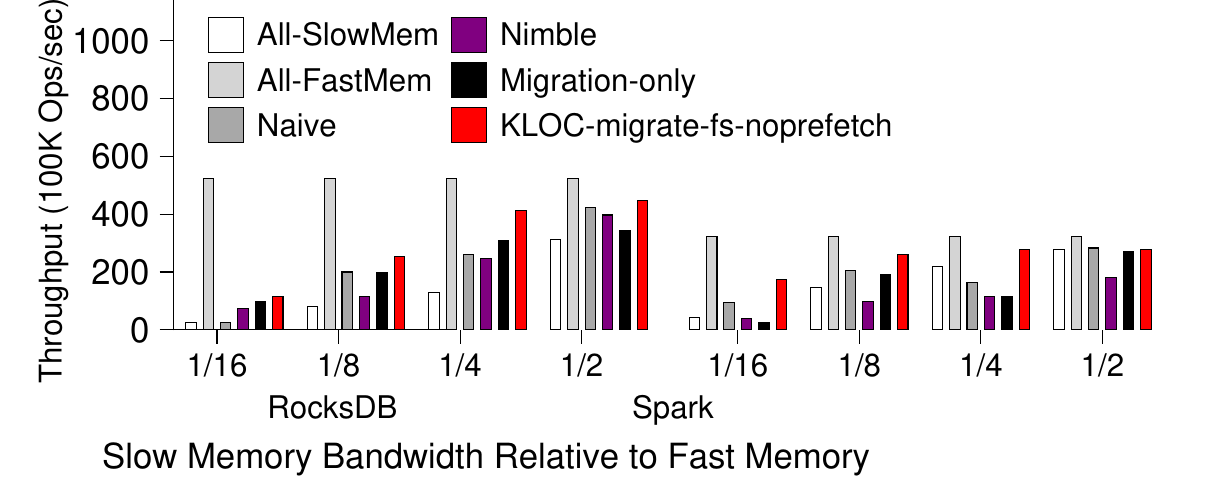}
  \captionsetup{width=0.95\linewidth}%
  \caption{{\bf Performance sensitivity to slow memory bandwidth.}
 {\footnotesize The x-axis varies slow memory's bandwidth ratio relative to
  fast memory, and the y-axis shows throughput. The fast memory capacity is set of
  5 GB (1/8th) of slow memory.}} 
  \label{fig:e-rocks-sensitivity-BW}
\end{subfigure}
  \vspace{-0.05in}
\caption{\bf Performance and memory bandwidth sensitivity.}
  \vspace{-0.1in}
\end{figure}

\subsection{Impact of \systemname}

\begin{figure}[t]
\centering
\includegraphics[width=180pt,height=120pt]{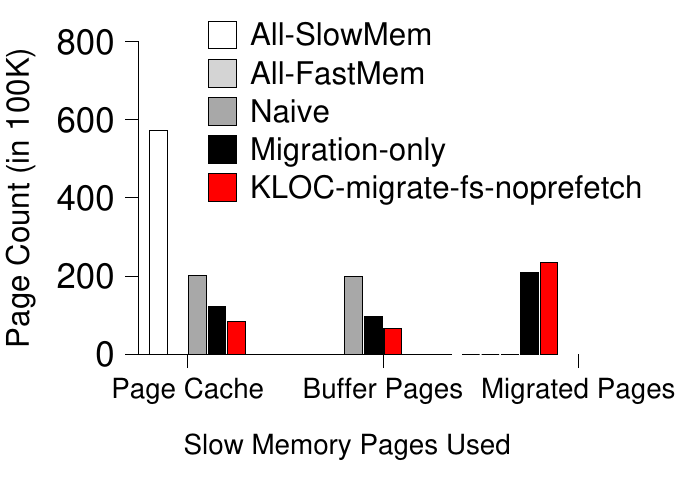}
 \vspace{-0.1in} 
 \caption{{\bf RocksDB's kernel page slow memory use.}  \small
  \footnotesize The x-axis shows slow memory pages allocated for  cache, kernel
  buffers (for OS data structures) or inactive pages migrated to slow memory.
  {\systemname} reduces slow memory use for page cache and kernel buffers.
  All-Fastmem does not use slow memory or incur migrations.}
 \label{fig:e-rocksdb-kernstat}
\end{figure}

We compare the following approaches summarized in Table~\ref{tab:sysmechanism}:
(1) All-SlowMem - the worst-case baseline, (2) All-FastMem (ideal case), (3)
Naive, (4) Migration-only, (5) {\systemname}-nomigrate -- the proposed
{\systemname} based placement of kernel objects related to file and socket
without migration, and finally, (6) {\systemname}-migrate-fs-noprefetch --
{\systemname}. To understand the reduction in the use of slow memory pages for
kernel objects (and hence better performance) with {\systemname}, in
Figure~\ref{fig:e-rocksdb-kernstat}, we show statistics for RocksDB. The x-axis
shows the total page cache, kernel buffer pages (slab pages for kernel in-memory
structures, logs, socket buffers, etc.), and inactive pages migrated to slower
memory, and the y-axis shows the page count (in units of 100K pages). 

\par\noindent{\bf Observations. }
First, as expected, incorrect placement of kernel-level objects to slow memory
can significantly degrade performance, as shown by the difference between the
All-FastMem (Optimal) vs. All-SlowMem configurations for all applications.
Second, the~\emph{Naive} greedy approach attempts to naively allocate~\emph{all}
application and OS pages to the limited-capacity fast memory. As a result, the
naive placement of non-critical kernel objects can drastically impact
performance-sensitive application pages and kernel objects. RocksDB uses
hundreds of 4~MB files to store the persistent key-values as a string-sorted
table~\cite{Kannan:NoveLSM, PebblesDB}; once the files are filled, low capacity
fast memory is polluted with inactive file caches and kernel objects (e.g.,
inode structures, dentry caches). As a result, the throughput of the
\emph{Naive} approach reduces by \RocksDBNaiveVsOptimal compared to the optimal
case. We observe a similar trend for Cassandra and Filebench. In contrast, Redis
uses few large files to checkpoint data, thereby even the\emph{ Naive} approach
provides gains over the worst-case baseline, but the cache page pollution
quickly hinders fast memory benefits. Spark, is composed of multiple
applications, which includes the Scala-based Spark
benchmark, the Spark framework, and the HDFS storage server (Hadoop file
system). While Spark by itself is compute-intensive, the HDFS storage server is
highly I/O intensive. As a result, the \emph{Naive} approach is ineffective.
\hl{Next, Nimble mainly tracks application-level pages using the OS swapping mechanism's active and inactive lists.  Lack of kernel page placement and excessive migration without a context hurts performance for I/O-intensive applications.}
\par Next, the \emph{Migration-only} approach only performs application and
kernel page migrations by tracking fast memory for cold (inactive) pages and
migrating them to slow memory. However, blindly migrating all kernel (and
application) objects increases page allocation and TLB and page table invalidation
overheads. Further, the \emph{Migration-only} approach lacks the semantic
information about how entities like files and sockets and their related kernel
objects are used. For example, even after an \textit{close()} operation, 
cache pages and journal data could continue to pollute faster memory.

Finally, the \emph{\systemname-nomigrate} approach provides effective placement
of kernel objects and file-related object and attempts to place them to fast
memory, but does not support page migration. In contrast,
\emph{{\systemname}-migrate-fs-noprefetch}, in addition to {\systemname}
placement, also aggressively demotes objects of inactive files to slower memory
using the modified LRU mechanism; because of this, the fast memory allocation
misses (or slow memory use) for active kernel objects also reduces(see
Figure~\ref{fig:e-rocksdb-kernstat}). As a result, RocksDB throughput improves
by up to \rocksnvmobjperf over migration approach. Interestingly, for Redis, the migration-based
\emph{{\systemname}-migrate-fs-noprefetch} approach provides up to 1.70x gains
even over \emph{\systemname-nomigrate}. Finally, the Java-based Cassandra uses a
large in-memory cache, and the application-level pages consume 95\% of fast
memory capacity, resulting in marginal gains with {\systemname}. 
\SK{Finally,  Spark, despite being compute-intensive, we observe up to
\sparkcontextgains gains with {\systemname}. These gains stem from (1) placement
of cache pages in fast memory accelerates Spark's checkpoint of storing large
RDD files to disk, and (2) removing inactive cache pages increases the use of
fast memory for application pages.}

\subsection{Sensitivity to Memory Bandwidth}
Figure~\ref{fig:e-rocks-sensitivity-BW} varies the slow memory bandwidth along
the x-axis and studies the impact on all RocksDB and Spark for brevity. The
x-axis varies slow memory's bandwidth ratio relative to fast memory, and the
y-axis shows throughput. The fast memory capacity is set of 8GB (1/10th) of slow
memory. Because RocksDB is highly memory intensive, with a significant fraction
of memory pages allocated in kernel and high kernel-level memory references, the
results show high performance benefit across all bandwidth configuration. For
Spark, \emph{{\systemname}-migrate-fs-noprefetch} benefits are high  until the
memory bandwidth is 1/4th of the fast memory. Increasing the slow memory
bandwidth further (e.g., 1/2 of fast memory) shows limited gains.  

{\textbf{Summary. } The results highlight the benefits of introducing file
context to encapsulate kernel objects and only target placement of objects currently
accessed by an application and avoiding placement cost of non-critical
objects.}

\begin{figure}[t]
\centering
  \vspace{-0.2in}
\includegraphics[width=7cm,height=4cm]{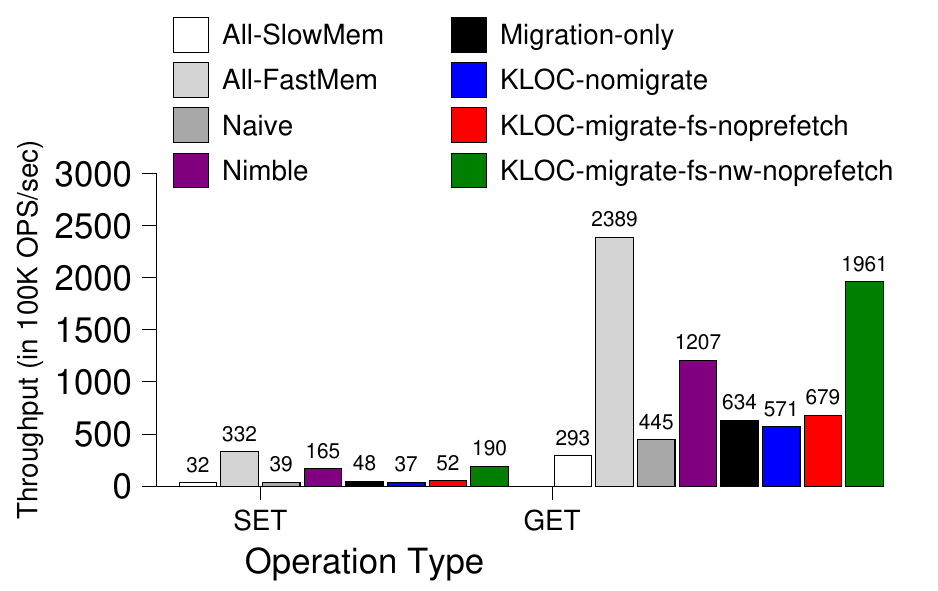}
  \vspace{-0.1in}
  \caption{{\bf Redis performance with {\systemname} for network stack.}
  {\footnotesize Results show cumulative throughput 8 Redis server instances that use 1~KB value size for 
  4-million keys with 75\% read (GET) requests.}}
\label{fig:e-redis}
\end{figure}

\begin{figure*}[t!]
\vspace{-0.25in}
\begin{minipage}[c][1\width]{0.33\textwidth}
  \centering
  \includegraphics[width=.90\linewidth]{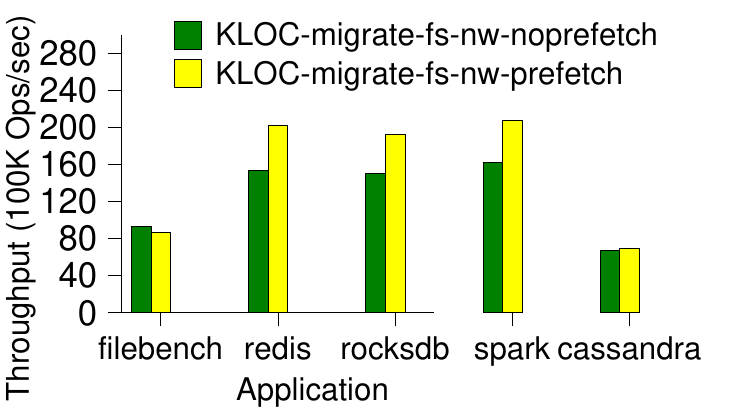}
  \captionsetup{width=0.95\linewidth}%
  \caption{{\bf {\systemname} I/O prefetch gains.} 
  {\footnotesize Y-axis is scaled application throughput (OPS/sec).}}
  \label{e-prefetch-allapp}
\end{minipage}
\begin{minipage}[c][1\width]{0.33\textwidth}
\centering
\footnotesize
\begin{tabular}{|p{1.5cm}|p{1.5cm}|p{1.5cm}|}
\hline
Access Pattern & KLOC-migrate-fs-nw-noprefetch & KLOC-migrate-fs-nw-prefetch\\ \hline
RandWrite & 53 & 55\\ \hline
RandRead   & 145 & 146\\ \hline
SeqWrite & 330 & 300\\ \hline
SeqRead  & 1178 & 1306\\ \hline
ReadWrite & 76 & 217\\ \hline
\end{tabular}
\captionsetup{width=.8\linewidth}
\caption{{\bf RocksDB throughput breakdown. }\footnotesize{Values show throughput in  (100K Ops/sec)}}
\label{fig:e-prefetch-rocksdb}
\end{minipage}
\begin{minipage}[c][1\width]{0.33\textwidth}
  \centering
  \includegraphics[width=6cm,height=3.5cm]{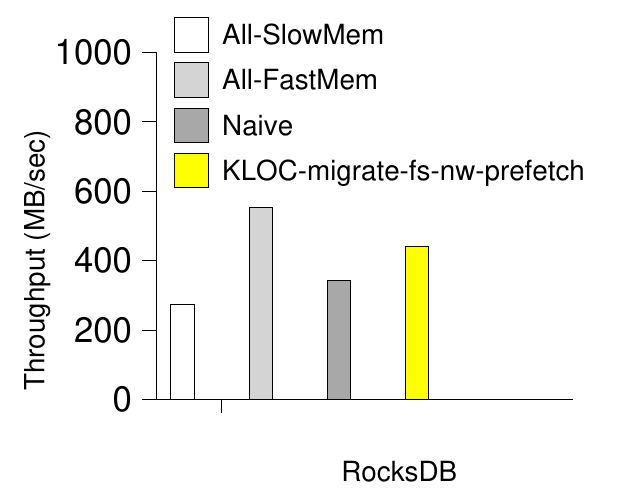}
  \captionsetup{width=.8\linewidth}
  \vspace{-0.2in}
  \caption{\bf {\systemname} impact on DC-Optane for RocksDB.} 
  \label{e-prefetch-allapp-dcoptane}
\end{minipage}
\vspace{-0.6in}
\end{figure*}

\subsection{Impact of {\systemname} for Network Stack}
We next discuss the benefits of introducing {\systemname} support for the socket
entity in the network stack and controlling the allocation and placement of
related kernel objects in heterogeneous memory. Figure~\ref{fig:e-redis} shows
the performance of Redis, a network-intensive key-value store serving hundreds
of clients. In contrast to the {\systemname} approach studied in
Figure~\ref{fig:e-allapps}  without the network stack support,
\emph{{\systemname}-migrate-fs-nw-noprefetch} in Figure~\ref{fig:e-redis} shows
the performance of network stack with {\systemname} support.  \par For
performance analysis, we use the well-known Redis benchmark performing 4 million
key operations, 25\%  insert (SET) and 75\%, fetch (GET) operations, and 1~KB
value size for each key as used by prior work~\cite{Gutierrez:integrated}.
Because Redis is a single-threaded application, we run \emph{8} instances of
Redis server with each instance using a set of dedicated ports. The y-axis shows
the throughput of SET and GET operations. Note that, for
\emph{{\systemname}-migrate-fs-nw-noprefetch}, each socket entity has an
independent {\systemname} map to encapsulate, allocate, and migrate kernel
objects and in addition to {\systemname}s for a file. As noted
in~\cref{sec:motivate} and~\cref{sec:design}, for the network subsystem, the
socket buffers (a.k.a \textit{skbuff}) dominate the kernel object allocation.
The socket buffers are allocated and reused across different layers (system
calls, TCP, IP, and NAPI) and can are reused across operations.

Redis is a memory-intensive application; the application's key-value store, the
network stack, and file system for checkpointing, all demand memory pages and
are sensitive to fast memory capacity and slow memory bandwidth. Compared to the
optimal case ( fast memory-only system), other approaches with limited faster
memory capacity suffer slowdown for both SET and GET operations. For the
\emph{Naive} approach, contention for fast memory across application and I/O
subsystems lead to incorrect placement and performance slowdown. The shorter
lifetime of network socket buffers, which are frequently allocated, released, or
reused across network operations, makes the \emph{Migration-only} approach
ineffective; this approach suffers from high page migration and related
overheads~\cite{Meswani:HeteroSW, Gupta:HetroVisor, Kannan:HeteroOS} reaping marginal benefits from memory heterogeneity. The
\emph{{\systemname}-migrate-fs-noprefetch} approach can only handle an efficient
kernel page placement of the file system's kernel object. In contrast,
\emph{{\systemname}-migrate-fs-nw-noprefetch} can map and efficiently handle the
placement of both network and file system kernel objects. The network supported
approach first attempts to allocate objects mapped to a socket's
\emph{\systemname} to  faster memory; when a direct allocation is not feasible,
it uses the modified Linux LRU-based migration approach to move inactive socket
related pages to slow memory and makes room for subsequent allocations to fast
memory. As a result, {\systemname} provides \eredisnetworkperf higher throughput
compared to the migration-based approach.

{\textbf{Summary. } The substantial performance gains highlight the benefits of
introducing a fine-grained {\systemname} support for socket to encapsulate
network stack objects for efficient data placement. Note that {\systemname} 
supports multiple instances (processes) highlighting the generality of
the proposed abstraction.}

\subsection{Exploiting I/O Stack Hints in \systemname}
\label{eval:prefetch}

Enabling {\systemname} to group set of kernel-level objects associated with
entities like files provides a capability to exploit OS-level hints and
optimizations for page placement. To showcase such flexibility, we evaluate the
benefits of combining {\systemname} and the filesystems' I/O prefetcher as shown
in Figure~\ref{e-prefetch-allapp}. The I/O prefetcher adapts to the application's I/O
access pattern and varies the page cache allocation behavior. In OSes such as
Linux, the I/O prefetcher expands the I/O prefetch window size (up to 128~MB)
adaptively for both spatial or temporal locality~\cite{He:Contract}. The page
prefetch window shrinks for random access patterns. Capturing such semantic
hints can be beneficial for {\systemname}'s page placement and reducing
migration. \SKREV{The naive baseline also uses I/O prefetcher but placing 
pre-fetched pages to fast memory depending on the space availability.} 

Figure~\ref{e-prefetch-allapp} shows the throughput of for all applications and
Figure~\ref{fig:e-prefetch-rocksdb} shows RocksDB's throughput for sequential
and random access I/O patterns (in the X-axis). To demonstrate the incremental
benefits, for brevity, we only compare {\systemname} without prefetcher support
(\emph{{\systemname}-migrate-fs-nw-noprefetch}) with prefetcher supported
\systemname (\emph{{\systemname}-migrate-fs-nw-prefetch}).

\SKREV{First, we observe that for naive, due to lack of context, the fast memory is either polluted 
with application pages or stale cache pages, reducing the placement of performance-critical I/O pages.}
Next, combining the prefetch I/O optimization in file system with {\systemname}
improves the performance of several I/O-intensive applications with temporal or
spatial locality. For example, applications such as RocksDB, Redis, and
Cassandra benefit from proactively placing prefetched I/O pages to faster
memory. For example, RocksDB's overall throughput improves
by~\rocksdbprefetch. As shown in Figure~\ref{fig:e-prefetch-rocksdb},
RocksDB's sequential access significantly benefits with prefetching. For
Filebench, we use random read and write workload that neither gains nor loses
performance; this is because, the prefetch approach uses the "I/O prefetch
window size" as a hint to predict random access and avoids aggressively placing
and polluting fast memory that are not likely to be used. As a side effect, this
reduces the migration of inactive fast memory pages to slow memory also
contributing towards performance benefits. For Redis, the performance gains
(~\redisprefetch) are from a periodic checkpoint of the in-memory key-value
store to the storage, which is mostly sequential. The overall Redis 
performance improves by 4$\times$ compared to migration-only approach.

\par {\bf Summary. } The results show the benefits of combining {\systemname} with
traditional OS-level optimizations such as I/O prefetcher. We believe the
techniques could be inherited for other subsystems~\cite{Yang:GPUPrefetch}.

\subsection{\systemname performance on DC-Optane Memory}
\vspace{-0.05in}
\label{eval:dcoptane}
\hl{Finally, to understand the impact of} {\systemname} \hl{ on DC-Optane memory
technologies, we use a 256GB DC-Optane on a memory socket with 16GB DRAM
hardware managed cache (memory mode) as a slow memory and a 48GB DRAM only
socket (fast memory). In addition to various limitations discussed in Section 2,
due to space constraints, we only show the results for RocksDB. First, as shown
in} Figure~\ref{e-prefetch-allapp-dcoptane}\hl{, only running RocksDB on slower
memory shows substantial performance degradation compared to the optimal (using
only fast memory) approach. Current DC-Optane technology in a memory mode
manages DRAM-cache as a direct-mapped cache. For large working set size, we see
a substantial increase in latency as well as throughput reduction, possibly
contributed due to a combination of poor cache management and high cache miss
overheads. Employing} {\systemname} \hl{by maximizing placement of kernel as
well as OS pages to faster memory accelerates performance considerably compared
to fully leaving it to the hardware and using a naive placement.}

\section{Conclusion}
\label{sec:conclude}
To provide efficient memory placement and management of kernel objects in
heterogeneous memory systems, we present the {\systemname}, a mechanism that
encapsulates kernel objects with fine-grained contexts with entities such as
files and sockets and provides efficient data placement and migration without
requiring expensive hotness scanning. Our results on real-world applications
such as RocksDB and Redis show up to 1.4$\times$  and $4\times$ higher
throughput compared migration-based techniques. Our future work will explore
supporting KLOC for other subsystems (e.g., GPU).

\bibliographystyle{IEEEtranS}
\bibliography{bib/p}

\end{document}